\documentclass[prc,preprint,
  showpacs,showkeys,lengthcheck,
  nofootinbib,tightenlines,twocolumn,notitlepage,
  preprintnumbers,superscriptaddress]{revtex4-1}

\usepackage[utf8]{inputenc}
\usepackage{newtxtext}
\usepackage[mathcal]{euscript}

\usepackage{graphicx}
\usepackage[usenames,dvipsnames]{xcolor}
\graphicspath{{figures/}}

\usepackage{hyperref}
\hypersetup{colorlinks=true,citecolor=blue,linkcolor=blue,urlcolor=blue}

\usepackage{diagbox}
\usepackage{booktabs}

\usepackage{amsmath,amssymb,amsfonts}
\usepackage{slashed}
\usepackage{bm}

\usepackage{titlesec}
\titlespacing{\section}{4pt}{10pt plus 4pt minus 2pt}{8pt plus 2pt minus 2pt}
\def\hs{\hspace}

\def\no{\nonumber}

\begin{document}

\title{
  Quark spin and orbital angular momentum from proton GPDs
}

\author{Adam Freese}
\email{afreese@anl.gov}
\address{Argonne National Laboratory, Lemont, Illinois 60439, USA}

\author{Ian C. Clo\"{e}t}
\email{icloet@anl.gov}
\address{Argonne National Laboratory, Lemont, Illinois 60439, USA}

\begin{abstract}
  We calculate the leading-twist helicity-dependent
  generalized parton distributions (GPDs) of the proton
  at finite skewness
  in the Nambu--Jona-Lasinio (NJL) model of quantum chromodynamics (QCD).
  From these (and previously calculated helicity-independent GPDs)
  we obtain the spin decomposition of the proton,
  including predictions for quark intrinsic spin
  and orbital angular momentum.
  The inclusion of multiple species of diquarks is found to have a significant
  effect on the flavor decomposition,
  and resolving the internal structure of these dynamical diquark correlations
  proves essential for the mechanical stability of the proton.
  At a scale of $Q^2=4\,$GeV$^2$ we find that the up and down quarks carry an
  intrinsic spin and orbital angular momentum of
  $S_u=0.534$, $S_d=-0.214$, $L_u=-0.189$, and $L_d=0.210$,
  whereas the gluons have a total angular momentum of $J_g=0.151$.
  The down quark is therefore found to carry almost no total angular momentum
  due to cancellations between spin and orbital contributions.
  Comparisons are made between these spin decomposition results and lattice
  QCD calculations.
\end{abstract}

\maketitle

\section{INTRODUCTION}
How the proton's spin is shared among its constituents is one of the most
pressing open questions in hadron physics.
Ever since the European Muon Collaboration found that the quarks' intrinsic
spin falls far short of saturating the proton's total spin~\cite{Ashman:1987hv},
various theoretical efforts have gone both into accounting for the remaining
spin, and into exploring the theoretical foundations for
decomposing the proton's spin.
For a review, see Ref.~\cite{Leader:2013jra}.

A prominent gauge-invariant decomposition of spin was proposed by
Ji~\cite{Ji:1996ek} using the flavor-separated gravitational form factors:
\begin{align}
  J_a = \frac{1}{2}\Big( A_a(0) + B_a(0) \Big)
  \label{eqn:ji}
  \,,
\end{align}
where $a=q,g$ are the quark and gluon contributions.
This allows the proton's spin to be decomposed into total contributions
from each parton type.
Since the total intrinsic spin of quarks is a gauge-invariant quantity,
one may decompose $J_q$ further into spin and orbital angular momentum,
giving a proton spin decomposition:
\begin{align}
  \frac{1}{2} = \sum_q \Big( S_q + L_q \Big) + J_g
  \,.
\end{align}
This is called the Ji spin decomposition.
A gauge-invariant decomposition of $J_g$ into intrinsic and
orbital angular momentum is not possible in this framework.

While alternative spin decompositions exist,
the Ji spin decomposition has the virtue of being calculable from
leading-twist generalized parton distributions
(GPDs)~\cite{Dittes:1988xz,Ji:1996nm,Diehl:2003ny}.
In particular, polynomiality sum rules~\cite{Ji:1998pc}
relate the Mellin moments of GPDs to
gravitational and axial form factors,
which when evaluated at $t=0$ give access to the total and spin angular
momentum of partons.
The GPDs are themselves of great contemporary interest because of their
relationship to spatial light cone distributions~\cite{Burkardt:2002hr},
the proton's mass decomposition~\cite{Lorce:2018egm,Hatta:2018sqd},
and cross sections for hard exclusive reactions such as deeply virtual Compton
scattering~\cite{Ji:1996nm,Radyushkin:1997ki}
that can be measured at facilities such as Jefferson Lab
and an Electron Ion Collider.

It is therefore important to perform calculations of the proton's
helicity-dependent
and helicity-independent leading-twist GPDs within a single framework
to make a unified set of predictions.
It is vital that any model calculation respect the symmetries and low-energy
dynamical properties of quantum chromodynamics (QCD).
Accordingly, we calculate the proton's helicity-dependent GPDs using the
Nambu--Jona-Lasinio (NJL) model of
QCD~\cite{Vogl:1991qt,Klevansky:1992qe,Hatsuda:1994pi},
an effective field theory that preserves all the global symmetries of QCD,
reproduces dynamical chiral symmetry breaking,
and can simulate aspects of confinement through use of proper time
regularization~\cite{Ebert:1996vx,Hellstern:1997nv,Cloet:2014rja}.
Moreover, the NJL model has previously been used to calculate the
helicity-independent proton GPDs~\cite{Freese:2019eww},
and because these calculations are symmetry-preserving the baryon number,
momentum, and angular momentum sum rules are automatically satisfied,
as are constraints such as polynomiality and correct support properties.

This paper is organized as follows.
In Sec.~\ref{sec:formalism}, we discuss the formalism used for calculating the
helicity-dependent proton GPDs.
In Sec.~\ref{sec:results}, we present the results for the GPDs and for the spin
decomposition they entail.
Finally, in Sec.~\ref{sec:summary} we present a summary and outlook.

\section{FORMALISM FOR CALCULATING PROTON GPDS}
\label{sec:formalism}

The formalism for calculating the proton GPDs has been laid out already in
Ref.~\cite{Freese:2019eww}.
However, we briefly review the formalism here, with additional elaborations
relevant to the helicity-dependent case.
The proton is considered as a bound state of three dressed
quarks. The bound state amplitude is found by solving the Faddeev equation,
which is dominated by configurations with two of the quarks
in a diquark correlation~\cite{Cahill:1988dx}.
In this work, we consider quark-diquark configurations specifically,
in particular configurations with isoscalar, Lorentz scalar and
isovector, axial vector diquarks.
More information about the proton bound state amplitude can be found in
Ref.~\cite{Cloet:2014rja}.

The proton's helicity-dependent GPDs are defined from the axial bilocal
lightcone correlator~\cite{Dittes:1988xz,Ji:1996nm,Diehl:2003ny}:
\begin{align}
  A_{\lambda'\lambda}^q
  &=
  \bar{u}(p',\lambda')
  \Big[
    \slashed{n}\gamma_5\, \widetilde{H}^q(x,\xi,t) \nonumber\\
    &\hspace*{25mm}
    +
    \frac{\gamma_5 (n\Delta)}{2M_N}\,
    \widetilde{E}^q(x,\xi,t)
    \Big]
  u(p,\lambda)
  \,,
\end{align}
where $P = \frac{1}{2}(p'+p)$, $\Delta = p'-p$, $\xi = -2(\Delta n)/(Pn)$,
$t=\Delta^2$, and $n$ is a lightlike vector defining the light front.
The GPDs are Lorentz-invariant functions of the three explicitly written
Lorentz-invariant arguments,
and also dependent on a renormalization scale $\mu$ not notated above.
In the NJL model calculation,
we take $\mu = M = 400$~MeV~\cite{Cloet:2005pp,Cloet:2007em,Cloet:2014rja}.

\begin{figure}
  \centering
  \includegraphics[width=\columnwidth]{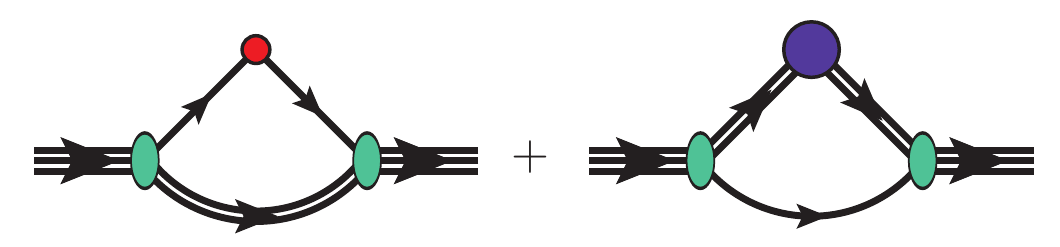}
  \caption{
    Diagrams contributing to the leading-twist proton GPDs.
    On the left is the quark diagram and on the right is the diquark diagram.
    The single line is the dressed quark propagator,
    the double line the diquark propagator, the shaded oval the Faddeev vertex,
    and the shaded circles represent the dressed quark and diquark GPDs.
  }
  \label{fig:proton:diagrams}
\end{figure}

The axial correlator itself is calculated by evaluating Feynman diagrams,
with the bilocal operator defining the GPDs inserted onto either a quark within
a diquark or on the accompanying quark,
both scenarios being depicted diagramatically in Fig.~\ref{fig:proton:diagrams}.
For the diquark propagators we implement the widely-used pole
approximation~\cite{Mineo:1999eq,Cloet:2005rt,Eichmann:2007nn,
Nicmorus:2008vb,Matevosyan:2011vj,Roberts:2011cf,Wilson:2011aa,
Segovia:2013uga,Carrillo-Serrano:2016igi}.
Self-consistency then demands that on-shell forms for the diquark GPDs
be used~\cite{Horikawa:2005dh},
even though they are in general off-shell. These approximations mean that the
inner structures of the diquarks are folded into the proton through
a convolution relation, which takes the form~\cite{Freese:2019eww}:
\begin{align}
  H_X(x,\xi,t)
  =
  \int \frac{\mathrm{d}y}{|y|}\,
  h_{Y/X}(y,\xi,t)\
  H_Y\!\left(\frac{x}{y},\frac{\xi}{y},t\right),
  \label{eqn:convolution}
\end{align}
where a hadron (proton) $X$ contains a composite hadron (diquark) $Y$,
and where $h_{Y/X}$ signifies ``body GPDs'' that encode the distribution
of $Y$ within $X$. The isospin weights for the quark and diquark diagrams,
for each quark flavor, are given in Eqs.~(102) and (103)
of Ref.~\cite{Cloet:2014rja}.

\subsection{Helicity-dependent diquark GPDs}
We proceed to consider the helicity-dependent GPDs of diquarks.
We first remark that scalar diquarks do not have helicity-dependent GPDs,
since the lack of total angular momentum does not provide a quantization axis.
Thus we need consider just axial vector diquarks
and transition GPDs between the two diquark species.

The axial vector diquark has four helicity-dependent GPDs.
We parametrize the on-shell correlator in the following way:
\begin{align}
  A_{a}^{q,\mu\nu}
  &=
  \frac{(n\Delta)}{\Delta^2}
  \frac{i\epsilon_{\Delta\mu \nu P}}{(Pn)}\ \widetilde{H}_{1a}^q(x,\xi,t) \no \\
&
  -
  \frac{
    i\epsilon_{n\Delta P \mu} \Delta^\nu
    - i\epsilon_{n\Delta P \nu} \Delta^\mu
    + i(P\Delta) \epsilon_{n\mu\nu\Delta}
  }{\Delta^2 (Pn)} \no \\
&\hs*{20mm} \times
  \left[
    - \widetilde{H}_{1a}^q(x,\xi,t)
    + \frac{\Delta^2}{M_a^2}\widetilde{H}_{2a}^q(x,\xi,t)
    \right]
  \no \\
  &- \frac{
    i\epsilon_{n\Delta P \mu} \Delta^\nu
    + i\epsilon_{n\Delta P \nu} \Delta^\mu
  }{M_a^2(Pn)}\
  \widetilde{H}_{3a}^q(x,\xi,t) \no \\
&
  + \frac{
    i\epsilon_{n\Delta P \mu} n^\nu
    + i\epsilon_{n\Delta P \nu} n^\mu
  }{2(P n)^2}\
  \widetilde{H}_{4a}^q(x,\xi,t)
  \label{eqn:gpd:axial:helicity}
  \,,
\end{align}
where $M_a$ is the axial vector diquark mass.
When contracted with polarization vectors $\varepsilon_\mu$
and $\varepsilon'^*_\nu$, this is equivalent to the standard
form given in Ref.~\cite{Berger:2001zb},
owing to a Schouten identity and the fact that the polarization
vectors are orthogonal to the diquark momenta.
We choose the form in Eq.~(\ref{eqn:gpd:axial:helicity})
in part because $\Delta$ has no virtuality dependence,
thus being preferred over $P$ for having a free Lorentz index,
and in part because it prevents the appearance of unphysical
poles in the axial form factors.
(See App.~\ref{sec:LT} for more details on the elimination
of these unphysical poles.)

Scalar-to-axial-vector and axial-vector-to-scalar (sa and as) transition GPDs
must be considered.
The bilocal axial correlator for scalar-to-axial transitions is:
\begin{align}
  A_{sa}^{q,\nu}
  &=
  \frac{n^\nu M_{as}}{(Pn)}
  \ \widetilde{H}_{sa,1}(x,\xi,t)
  + \frac{(n\Delta)\Delta^\nu}{(Pn)M_{as}}\
  \widetilde{H}_{sa,2}(x,\xi,t)
  \,,
\end{align}
where $M_{as} = M_s + M_a$ and $M_s$ is the scalar diquark mass.
There is an analogous expression for the axial-vector-to-scalar transition case.
These GPDs have the property of being neither T-even nor T-odd.
However, they remain related by time-reversal symmetry in a vital respect:
\begin{align}
  \widetilde{H}_{sa,i}(x,\xi,t)
  =
  - \widetilde{H}_{as,i}(x,-\xi,t)
  \,.
\end{align}
Crucially, the proton body GPDs accompanying these diquark GPDs in the
convolution formula Eq.~(\ref{eqn:convolution}) exhibit this same property,
which ensures that any T-odd contributions to the proton GPDs resulting from
the diquark transition diagrams cancel out---a necessity, since proton GPDs are
strictly T-even.

\subsection{Dressed quark GPDs}
\label{sec:dressing}
The dressed quarks in the NJL model are quasi-particles arising from an
amalgamation of nearly massless current quarks.
Since GPDs are defined using bilocal operators of current quark fields,
the dressed quarks have nontrivial GPDs that must be calculated within
the NJL model and folded into hadrons via Eq.~(\ref{eqn:convolution}).
As discussed in Ref.~\cite{Freese:2019eww},
the leading-twist dressed quark GPDs can be obtained by solving an
inhomogeneous Bethe-Salpeter equation.
For the helicity-dependent GPDs,
one has $\slashed{n}\gamma_5\delta(n[xP-k])$
as a driving term.

We find the isoscalar and isovector helicity-dependent
dressed quark GPDs to be:\footnote{
  These results are without $\pi$-$a_1$ mixing for consistency
  with Ref.~\cite{Cloet:2014rja}, from which we lift the model parameters.
}
\begin{subequations}
\begin{align}
\label{eqn:gpd:HtQ}
\widetilde{H}_{I=0,1}(x,\xi,t) &=\delta(1-x)\,, \\
\label{eqn:gpd:EtQ}
\widetilde{E}_{I=0,1}(x,\xi,t) &=
    \frac{N_c}{2\pi^2}
    \frac{1}{|\xi|}
    \frac{
      G_{\eta,\pi} M^2
    }{1 + 2G_{\eta,\pi} \Pi_{PP}(t)} \no \\
&\hs*{-1mm}\times
      \Gamma\left(0,\alpha/\Lambda_{\mathrm{UV}}^2,
      \alpha/\Lambda_{\mathrm{IR}}^2
      \right)\ \Theta(|\xi|-|x|)\,,
\end{align}
\label{eqn:gpd:quark}%
\end{subequations}
where $\Gamma(s,a,b) = \int_a^b \mathrm{d}t\,t^{s-1}\,e^{-t}$
is the generalized incomplete gamma function and
$\alpha = M^2- \frac{1}{4}(1-x^2/\xi^2)t$.
We remark that the support region for the dressing functions
[i.e., for the contributions to the GPDs other than $\delta(1-x)$]
is entirely constrained to the ERBL region,
and thus that the PDF in particular is undressed.
It's also worth noting that the GPD $\widetilde{E}_I(x,\xi,t)$ contains
a pion pole or $\eta$ meson pole, depending on the isospin.

\section{RESULTS}
\label{sec:results}

\begin{figure}
  \includegraphics[width=0.49\columnwidth]{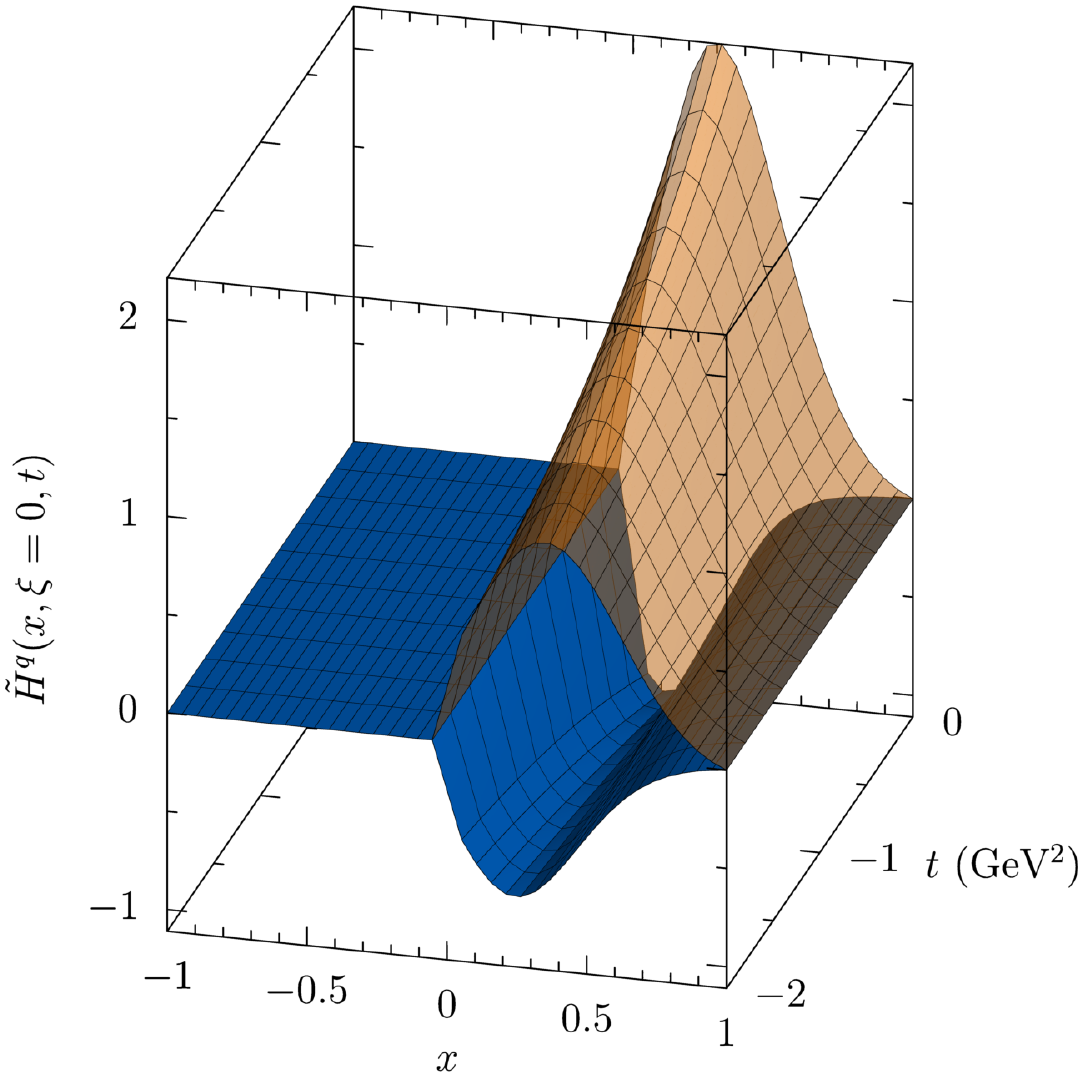}
  \includegraphics[width=0.49\columnwidth]{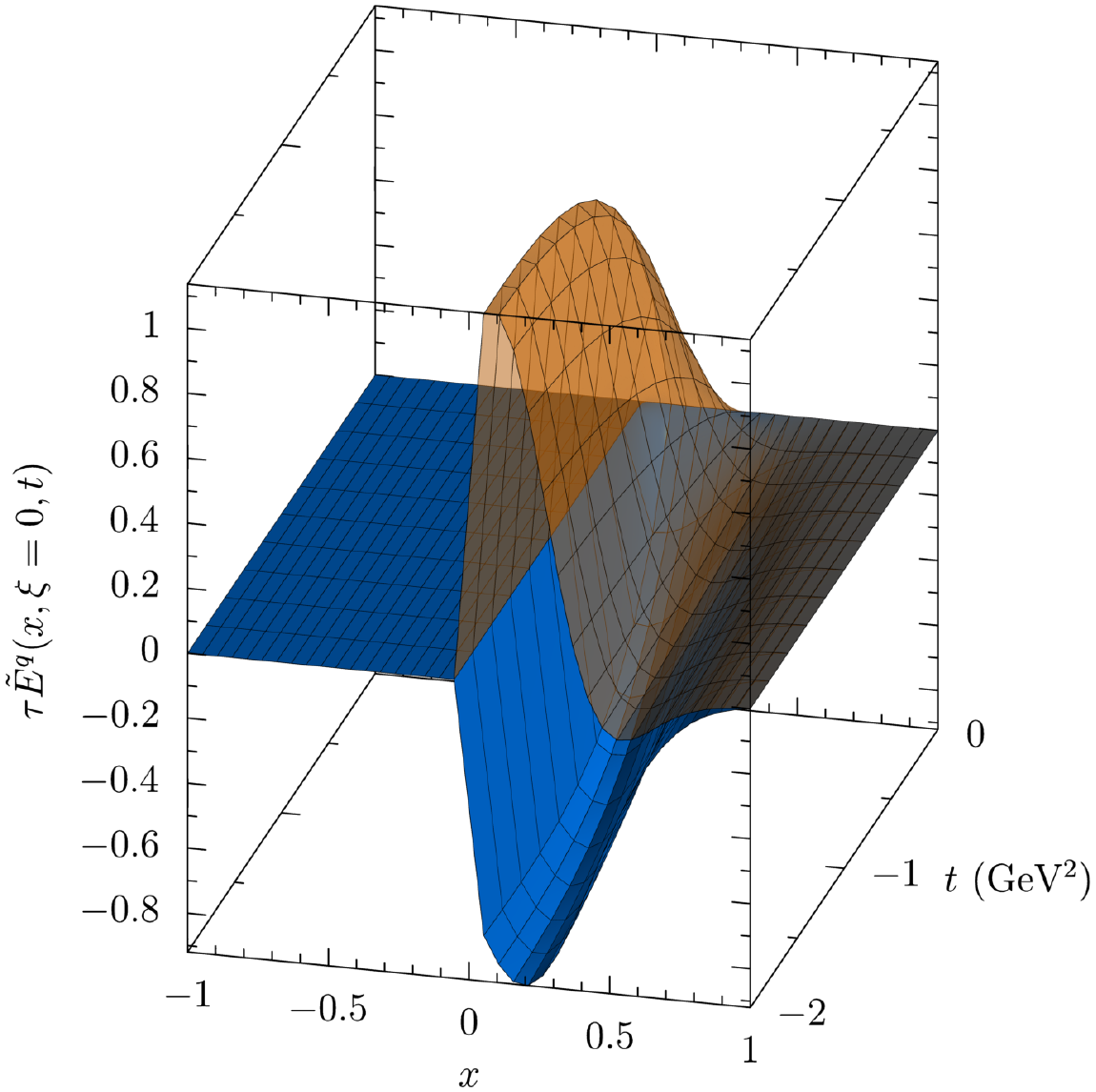} \\[0.5em]
  \includegraphics[width=0.49\columnwidth]{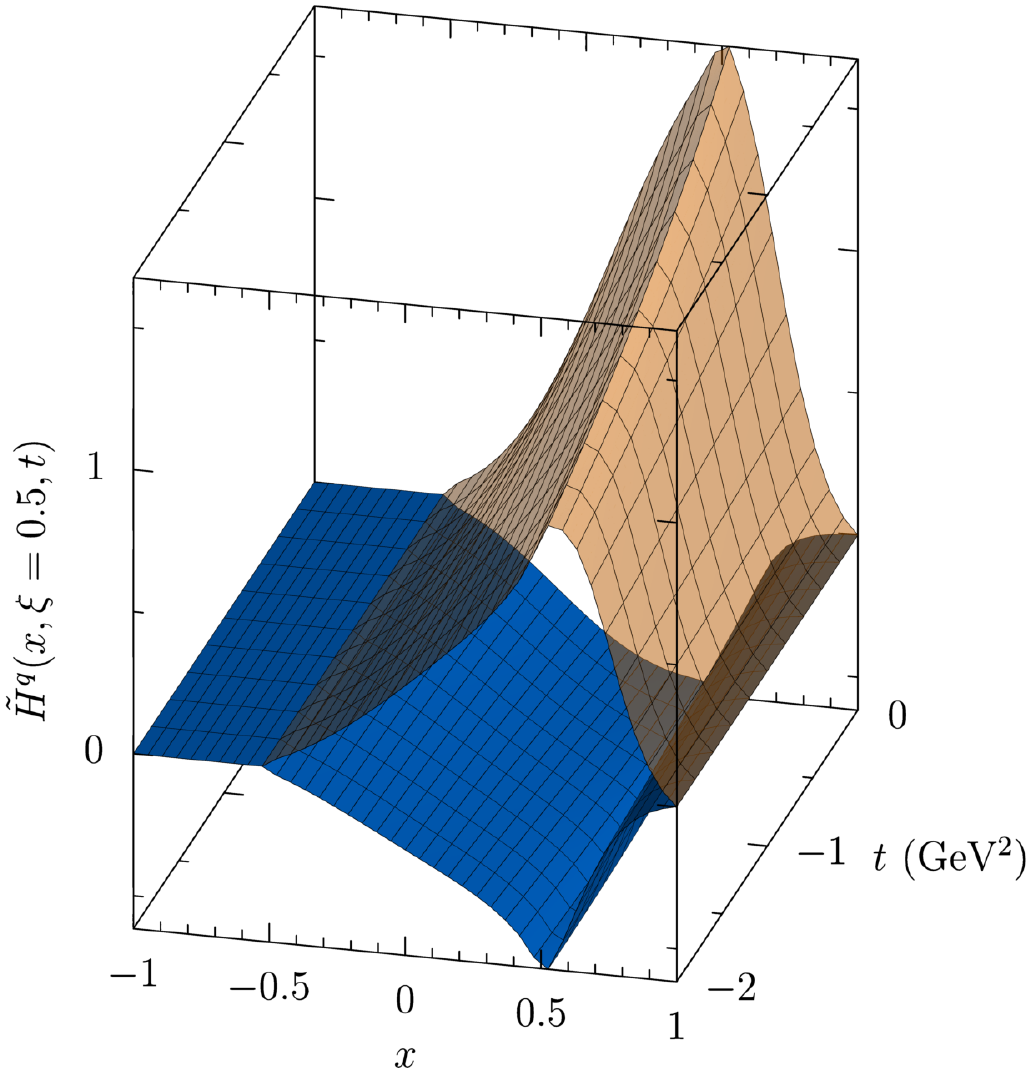}
  \includegraphics[width=0.49\columnwidth]{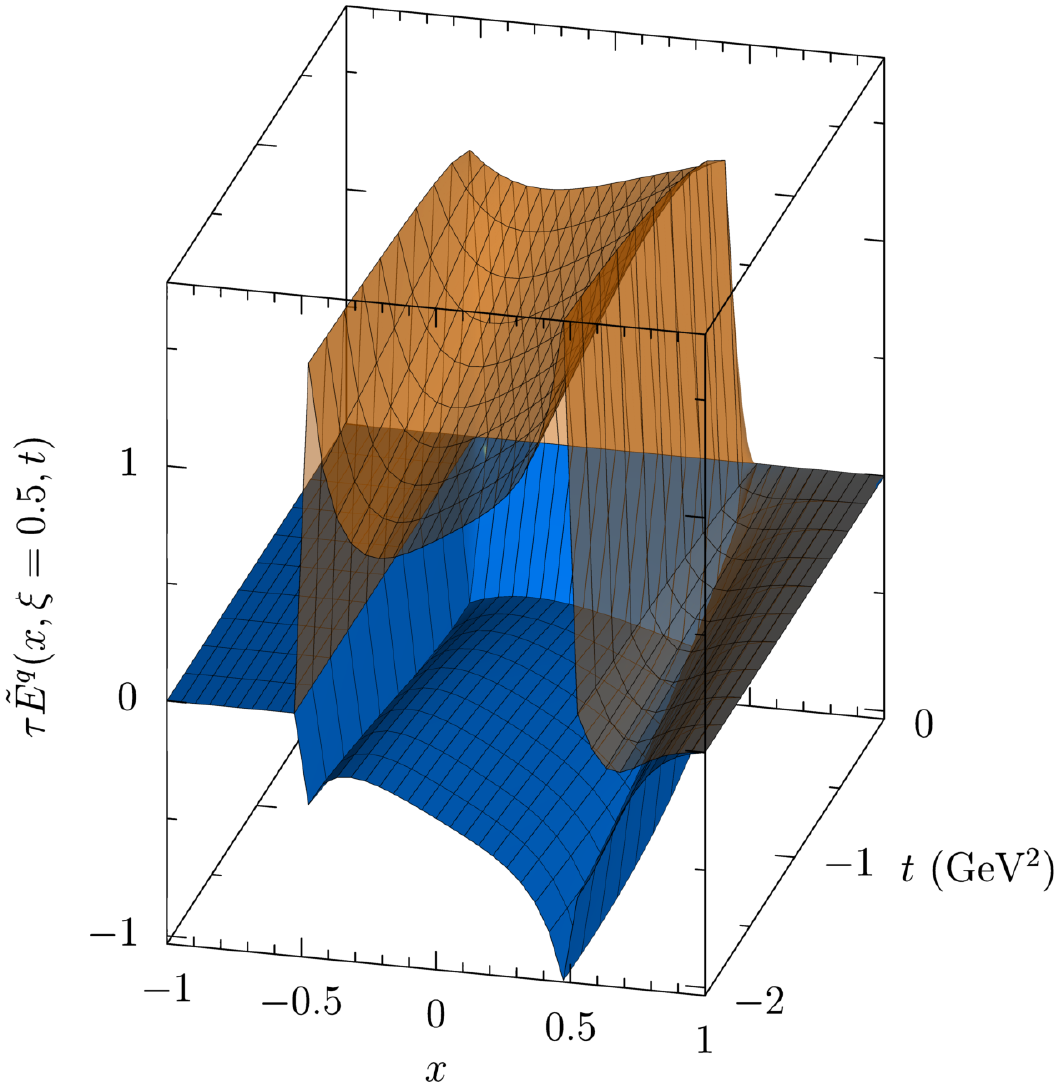}
  \caption{
    The helicity-dependent proton GPDs at the model scale
    of $Q^2=0.16\,$GeV$^2$,
    where the GPD $\widetilde{E}^q(x,\xi,t)$ has been scaled by a factor
    $\tau = -t/(4M_N^2)$.
    The top row is for $\xi=0$ and the bottom row has $\xi=0.5$.
    The transparent (orange) surface is up quarks and the opaque (blue)
    surface is down quarks.
  }
\label{fig:gpd:hel}
\end{figure}

With the formalism above, we proceed to present results for the
helicity-dependent GPDs of the proton,
as well as for the proton spin decomposition.
Specifically, the model parameters from Ref.~\cite{Cloet:2014rja} are used.
However, in addition, we also consider a model variant with only scalar
diquarks.
For this, the scalar diquark parameter $G_s$ is found by solving
proton's Faddeev equation with the proton mass fixed to its physical value.
In the scalar-only model, we find $G_s = 9.98$~GeV$^{-2}$ and $M_s = 576$~MeV.

\subsection{Helicity-dependent proton GPDs}

\begin{figure}
  \includegraphics[width=0.49\columnwidth,height=43mm]{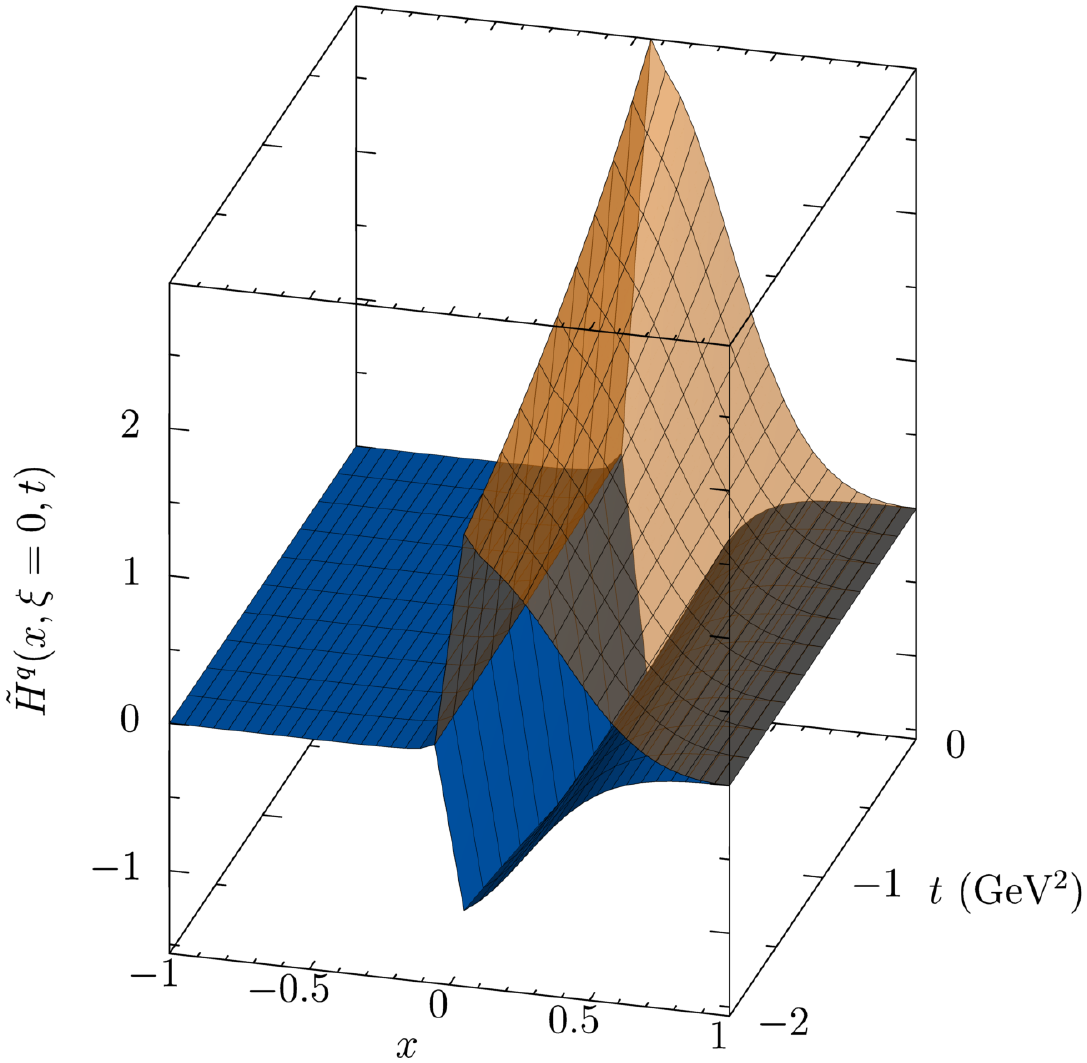}
  \includegraphics[width=0.49\columnwidth,height=43mm]{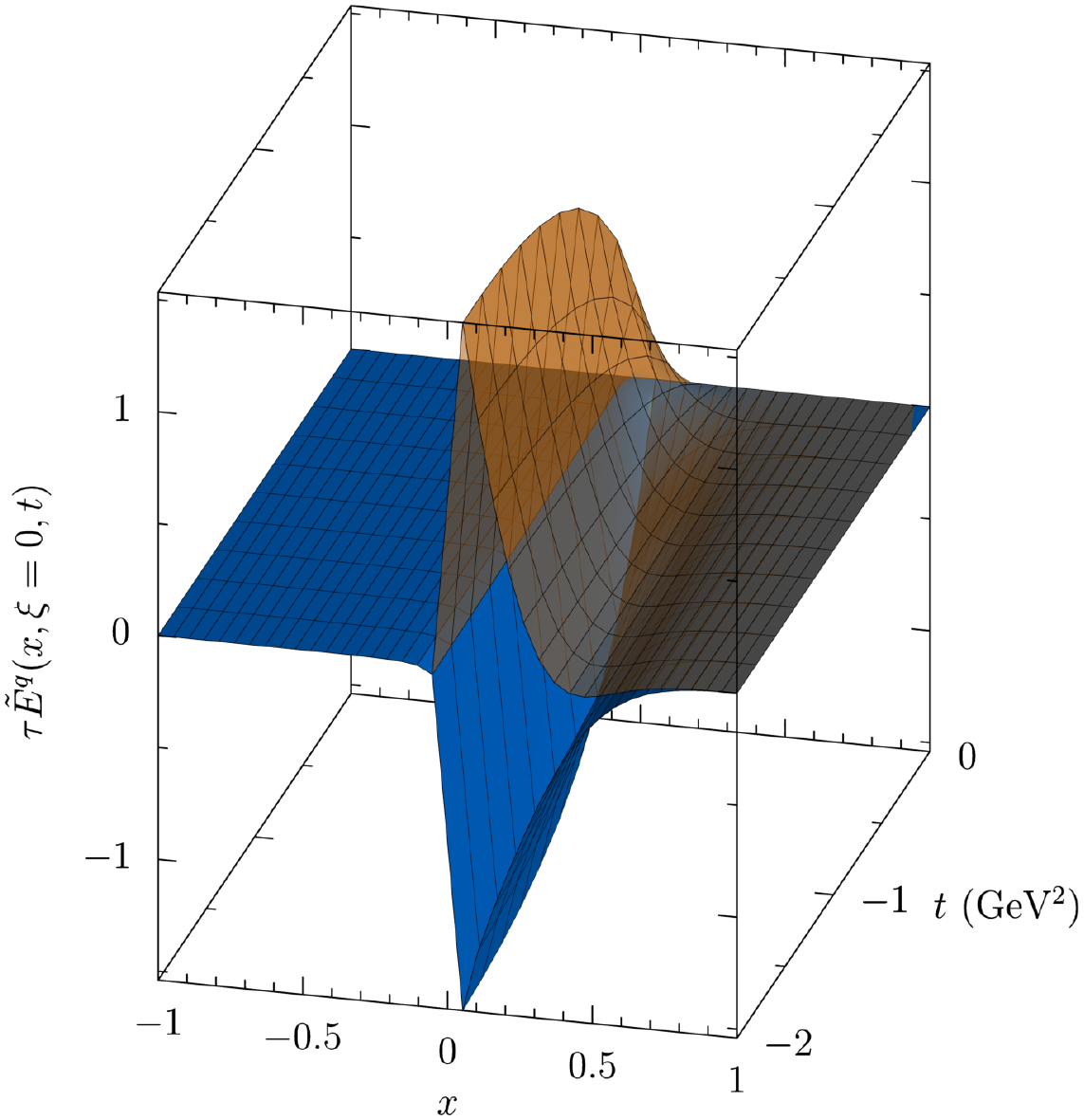} \\[0.5em]
  \includegraphics[width=0.49\columnwidth,height=43mm]{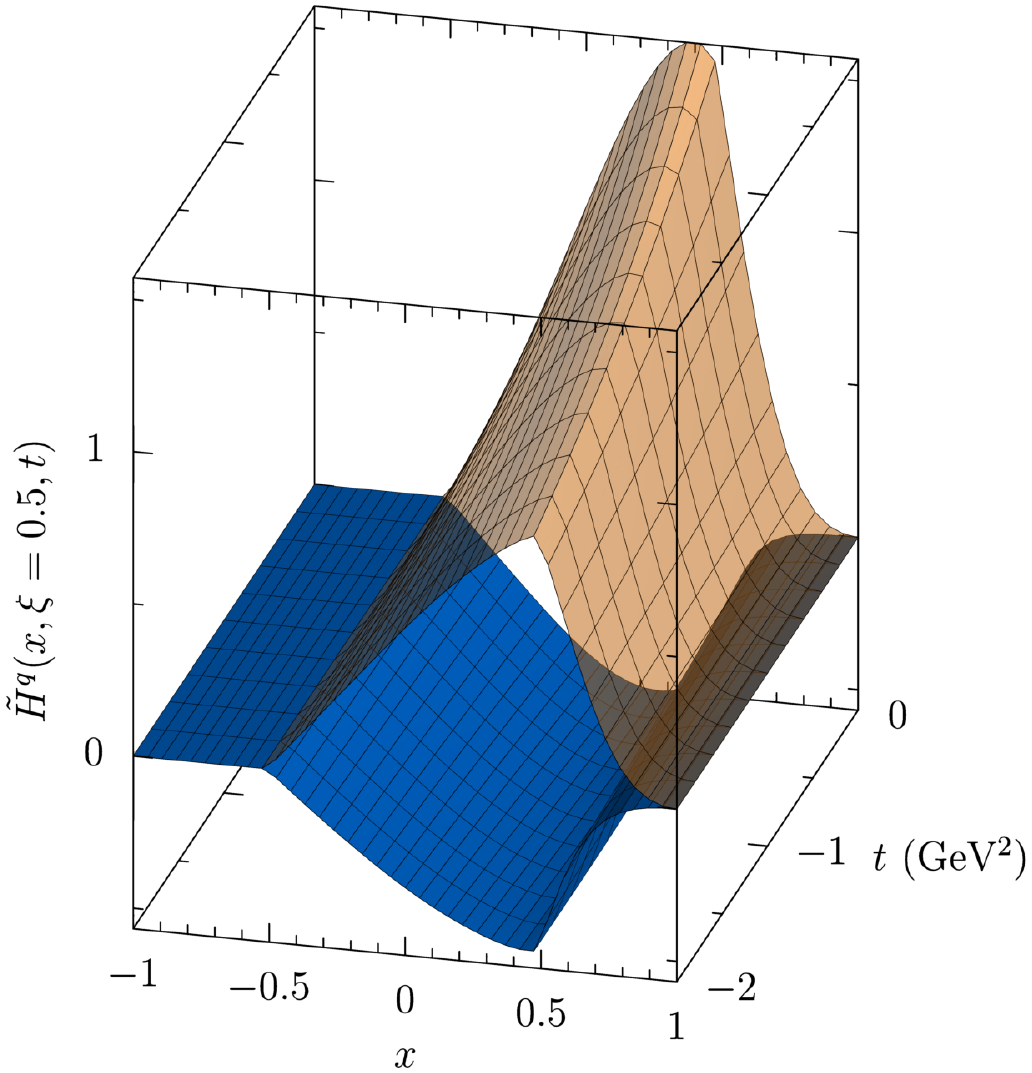}
  \includegraphics[width=0.49\columnwidth,height=43mm]{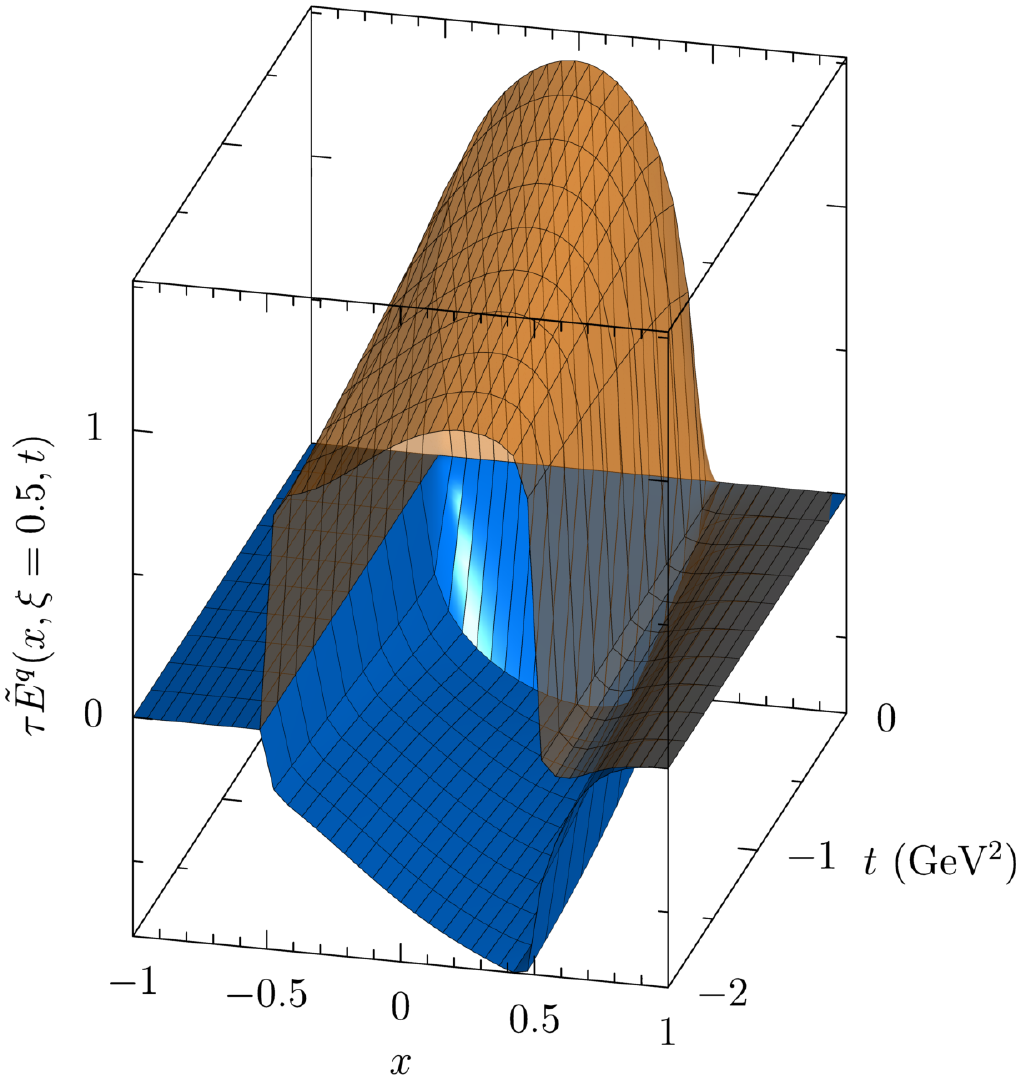}
  \caption{
    The helicity-dependent proton GPDs at the scale of $Q^2=4.0\,$GeV$^2$,
    where the GPD $\widetilde{E}^q(x,\xi,t)$ has been scaled by a factor
    $\tau = -t/(4M_N^2)$.
    The top row is for $\xi=0$ and the bottom row has $\xi=0.5$.
    The transparent (orange) surface is up quarks and the opaque (blue) surface
    is down quarks.
  }
\label{fig:gpd:hel:evo}
\end{figure}

Helicity-dependent GPDs are presented for zero and finite skewness ($\xi=0.5$)
at the model scale $Q^2 = M^2$ in Fig.~\ref{fig:gpd:hel}.
Since the helicity-dependent GPD $\widetilde{E}^q(x,\xi,t)$
becomes large near $t=0$ due to the presence of a pion pole,
it is scaled by a factor $\tau = -t/(4M_N^2)$.
We see that for $\xi=0$ our GPD results have no support for $-1< x < 0$ because,
at the model scale, we have not included anti-quarks in the model calculation.
However, at finite skewness an ERBL region ($-\xi < x < \xi$)
develops and our GPDs are non-zero in the range $-\xi < x < 1$,
even in this valence quark picture at the model scale.
These results clearly show that GPDs at finite skewness can display
radically different features from those at $\xi = 0$.

\begin{table*}[t]
  \setlength{\tabcolsep}{0.55em}
  \renewcommand{\arraystretch}{1.3}
  \caption{
    Decomposition of static properties of the proton by
    the various diagrammatic contributions,
    where the full scalar+axial diquark model is used.
    The quantities are given at the model scale of $Q^2 = 0.16\,$GeV$^2$
    and the diquark in the brackets is the spectator.
  }
  \label{tab:static:diagrams}
  \begin{tabular}{@{}lcccccccccccc@{}}
    \toprule
    Diagram &
    $ \kappa_u $ &
    $ \kappa_d $ &
    $ A_u(0) $ &
    $ A_d(0) $ &
    $ B_u(0) $ &
    $ B_d(0) $ &
    $ C_u(0) $ &
    $ C_d(0) $ &
    $ S_{\mathrm{tot}} $ &
    $ L_{\mathrm{tot}} $ &
    $ J_{\mathrm{tot}} $ &
    $ g_A $ \\
    \hline
    Quark (scalar) &
    $\phantom{-}1.134 $ &
    $ 0 $ &
    $ 0.248 $ &
    $ 0 $ &
    $\phantom{-}0.306 $ &
    $ 0 $ &
    $\phantom{-}0.020 $ &
    $ 0 $ &
    $\phantom{-}0.287 $ &
    $-0.100 $ &
    $\phantom{-}0.277 $ &
    $ 0.574 $ \\
    Scalar diquark &
    $-0.546 $ &
    $-0.546 $ &
    $ 0.220 $ &
    $ 0.220 $ &
    $-0.153 $ &
    $-0.153 $ &
    $-0.516 $ &
    $-0.516 $ &
    $ 0 $ &
    $\phantom{-}0.068 $ &
    $\phantom{-}0.068 $ &
    $ 0 $ \\
    Quark (axial) &
    $-0.150 $ &
    $-0.300 $ &
    $ 0.034 $ &
    $ 0.067 $ &
    $-0.060 $ &
    $-0.120 $ &
    $\phantom{-}0.039 $ &
    $\phantom{-}0.048 $ &
    $-0.066 $ &
    $\phantom{-}0.026 $ &
    $-0.040 $ &
    $ 0.044 $ \\
    Axial diquark &
    $\phantom{-}0.785 $ &
    $\phantom{-}0.157 $ &
    $ 0.176 $ &
    $ 0.035 $ &
    $\phantom{-}0.150 $ &
    $\phantom{-}0.030 $ &
    $-0.155 $ &
    $-0.031 $ &
    $\phantom{-}0.137 $ &
    $\phantom{-}0.058 $ &
    $\phantom{-}0.195 $ &
    $ 0.182 $ \\
    Transition diquark &
    $\phantom{-}0.346 $ &
    $-0.346 $ &
    $ 0 $ &
    $ 0 $ &
    $\phantom{-}0.108 $ &
    $-0.108 $ &
    $\phantom{-}0.014 $ &
    $-0.014 $ &
    $ 0 $ &
    $ 0 $ &
    $ 0 $ &
    $ 0.751 $ \\
    \hline
    Sum &
    $\phantom{-}1.569 $ &
    $-1.045 $ &
    $ 0.678 $ &
    $ 0.322 $ &
    $\phantom{-}0.351 $ &
    $-0.351 $ &
    $-0.598 $ &
    $-0.483 $ &
    $\phantom{-}0.358 $ &
    $\phantom{-}0.142 $ &
    $\phantom{-}0.500 $ &
    $ 1.551 $ \\
    \bottomrule
  \end{tabular}
\end{table*}

A visually significant aspect of the $\widetilde{E}^q(x,\xi,t)$
results in Fig.~\ref{fig:gpd:hel} is the jump discontinuities at $x=\pm\xi$.
This occurs in effective theories with a four-fermion interaction
vertex~\cite{Petrov:1998kf,Polyakov:1999gs,Theussl:2002xp},
and can be seen in the dressed quark GPD of Eq.~(\ref{eqn:gpd:EtQ}).
On the surface this is an apparent problem for QCD factorization,
which requires GPDs to be continuous across the DGLAP-ERBL boundary.
However, these jump discontinuities are removed by GPD evolution,
rendering the model calculations compatible with QCD factorization
above the model scale
and allowing Compton form factors to be rigorously calculated.

In Fig.~\ref{fig:gpd:hel:evo}, we present the same
helicity-dependent proton GPDs as in Fig.~\ref{fig:gpd:hel},
but evolved to a scale $Q^2 = 4\,$GeV$^2$
using leading-order kernels~\cite{Ji:1996nm,Radyushkin:1997ki,Vinnikov:2006xw}.
We find that the QCD evolution has a dramatic impact on
$\widetilde{E}^q(x,\xi,t)$,
which is now also continuous across the DGLAP-ERBL boundary.

With both the helicity-dependent proton GPDs above and the previously calculated
helicity-independent GPDs~\cite{Freese:2019eww} in hand,
we will proceed to consider various static properties of the proton,
with a special focus on its spin decomposition.

\subsection{Static properties of the proton}
\label{sec:static}

Various static properties of the proton can be obtained from Mellin moments
of the GPDs at $t=0$.
Several of these, such as the electric charge, magnetic moment,
axial charge, and quark spin $S_q$
can be obtained from form factors and have been studied elsewhere
(see Ref.~\cite{Cloet:2014rja} for electromagnetic properties).
Others, such as the total angular momentum $J$,
the anomalous gravitomagnetic moment $B(0)$,
and the D-term $C(0)$
are new opportunities afforded through GPDs.
The gravitational form factors $A(t)$, $B(t)$, and $C(t)$ can be obtained
from the helicity-independent GPDs through:
\begin{align}
  \sum_{a=q,g}
  \int_{-1}^1 \mathrm{d}x\,
  x H^a(x,\xi,t)
  &=
  A(t) + \xi^2 C(t)
  \\
  \sum_{a=q,g}
  \int_{-1}^1 \mathrm{d}x\,
  x E^a(x,\xi,t)
  &=
  B(t) - \xi^2 C(t)
  \,,
\end{align}
and the total angular momentum can then be obtained through the
Ji sum rule in Eq.~(\ref{eqn:ji}).
Moreover, by not summing over parton flavors,
one can obtain a flavor decomposition of these quantities,
although such a breakdown will be renormalization scheme and scale dependent
(unlike the sum, which is scheme and scale independent).

The quark spin can be obtained from the helicity-dependent GPDs:
\begin{align}
  S_q = \frac{1}{2} \int_{-1}^1 \mathrm{d}x \,
  \widetilde{H}^q(x,\xi,t=0)
  \,,
\end{align}
and the quark orbital angular momentum can then be obtained through
$L_q = J_q - S_q$.
The isovector axial vector charge $g_A$ is related to the up and down
intrinsic spin via the Bjorken sum rule: $g_A = 2(S_u - S_d)$.

We present the results for various static quantities of the proton,
along with a diagram-by-diagram breakdown,
in Tab.~\ref{tab:static:diagrams}.
In particular, these quantities are calculated with both scalar and
axial vector diquarks present in the proton.
The first two columns of results provide contributions to the proton's
flavor-separated anomalous magnetic moment,
which are included to provide a comparison with other results
and because they would vanish in the absence of
orbital angular momentum in the proton.
The next two columns provide quark momentum factors in the proton,
and we find that scalar diquark configurations carry about twice
the light-cone momentum as the axial vector configurations.
In addition, up quarks carry about two-thirds and down quarks about
one-third of the total light-cone momentum, as naively expected.

For the flavor separated quantities in Tab.~\ref{tab:static:diagrams}
we first remark that not only does the total $B(0)$ vanish
(as expected by angular momentum conservation)
but that the total contribution from each diquark configuration also vanishes.
However, this is not the case for $B_u(0)$ or $B_d(0)$ separately.
This is a similar observation to that found in Ref.~\cite{Brodsky:2000ii},
where each state in a Fock space expansion has $B(0)=0$,
and has the same formal cause:
the diquark configuration (or the Fock state) has the same quantum numbers
as the proton, and is thus a $J=\frac{1}{2}$ eigenstate. Thus we can say
$\langle J\rangle = \frac{1}{2}\langle x\rangle = \frac{1}{2}A(0)$
for each configuration (or Fock state) individually, entailing $B(0)=0$.

We next remark on the $C(0)$ contributions of the various diagrams.
The negativity condition~\cite{Perevalova:2016dln},
which states that $C(0) < 0$ is necessary for mechanical stability,
is satisfied by both diquark configurations.
In both cases, $C(0)$ is positive for the quark diagram and negative
for the diquark diagram.
This illustrates the necessity of resolving the dynamical diquark
degrees of freedom in order to obtain a mechanically stable proton.

\begin{figure*}[tbp]
  \includegraphics[width=0.31\textwidth]{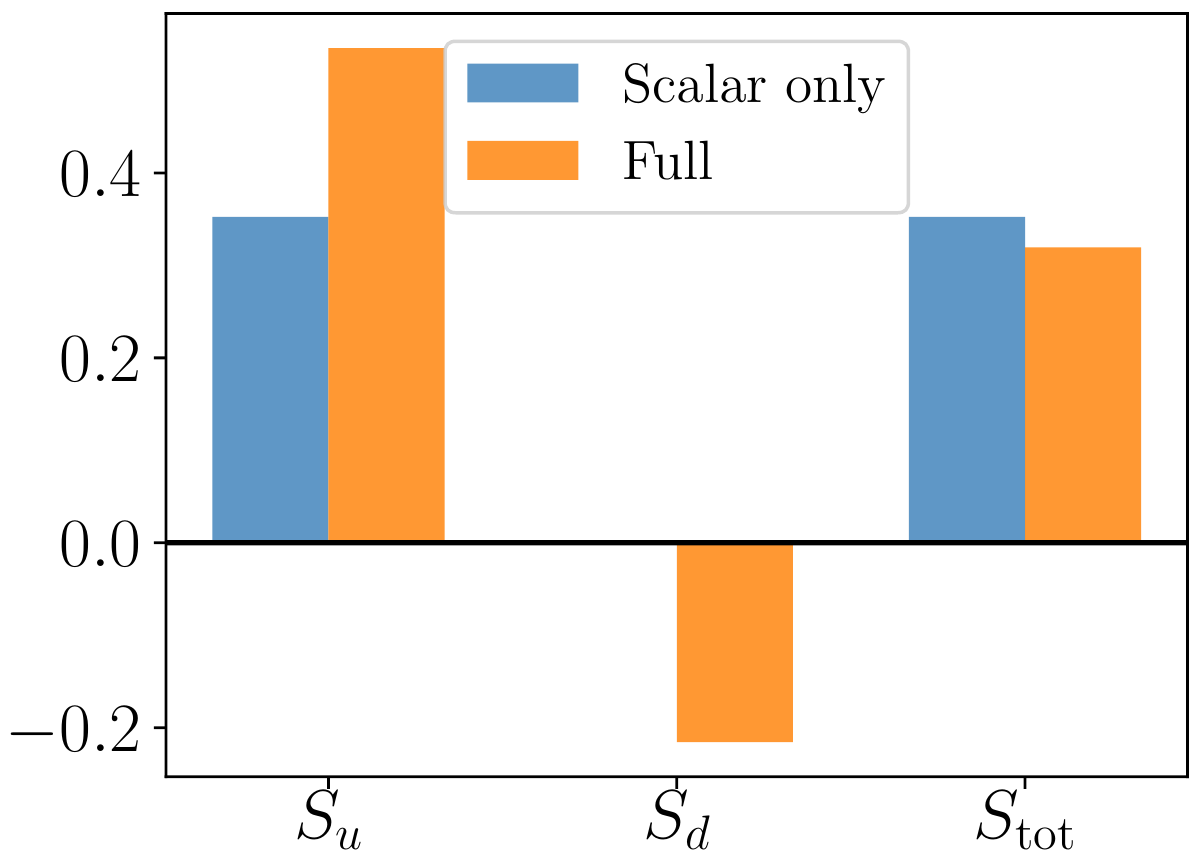} \hs*{3mm}
  \includegraphics[width=0.31\textwidth]{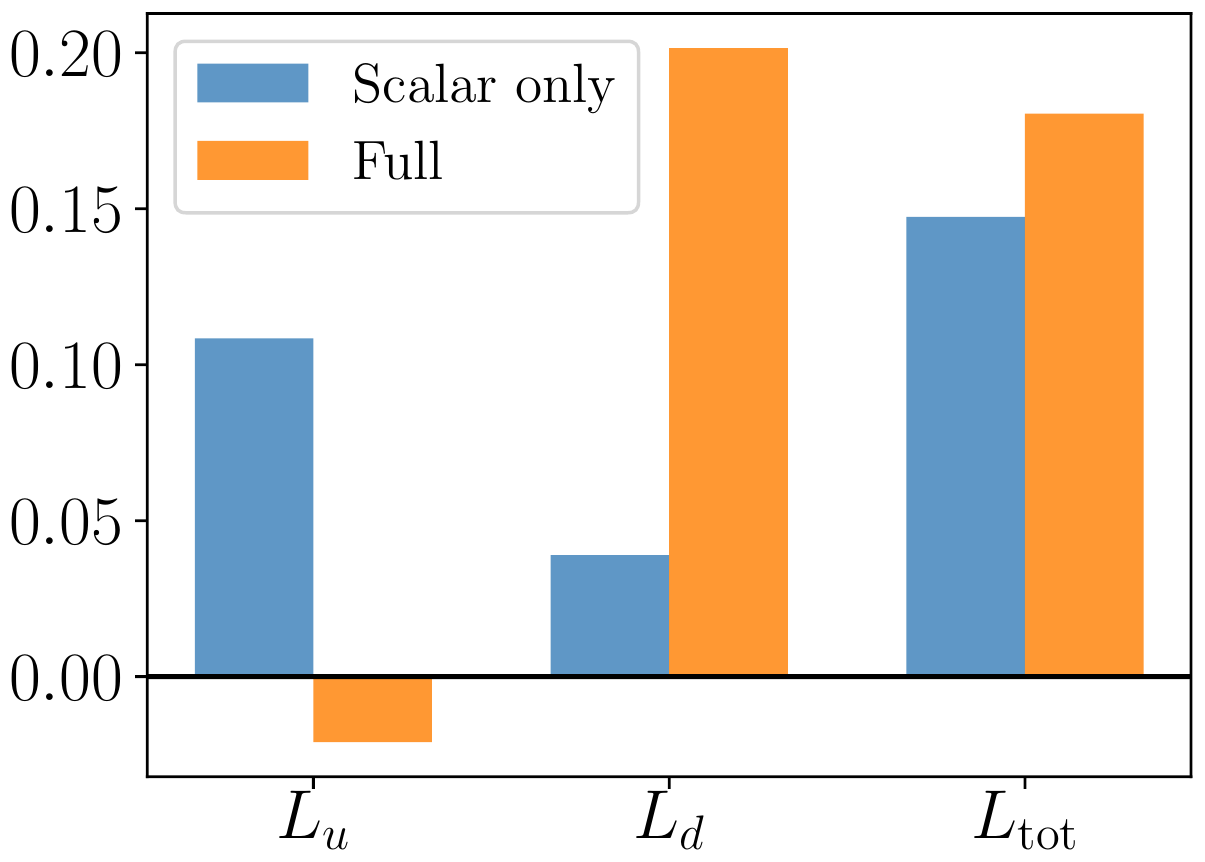} \hs*{3mm}
  \includegraphics[width=0.31\textwidth]{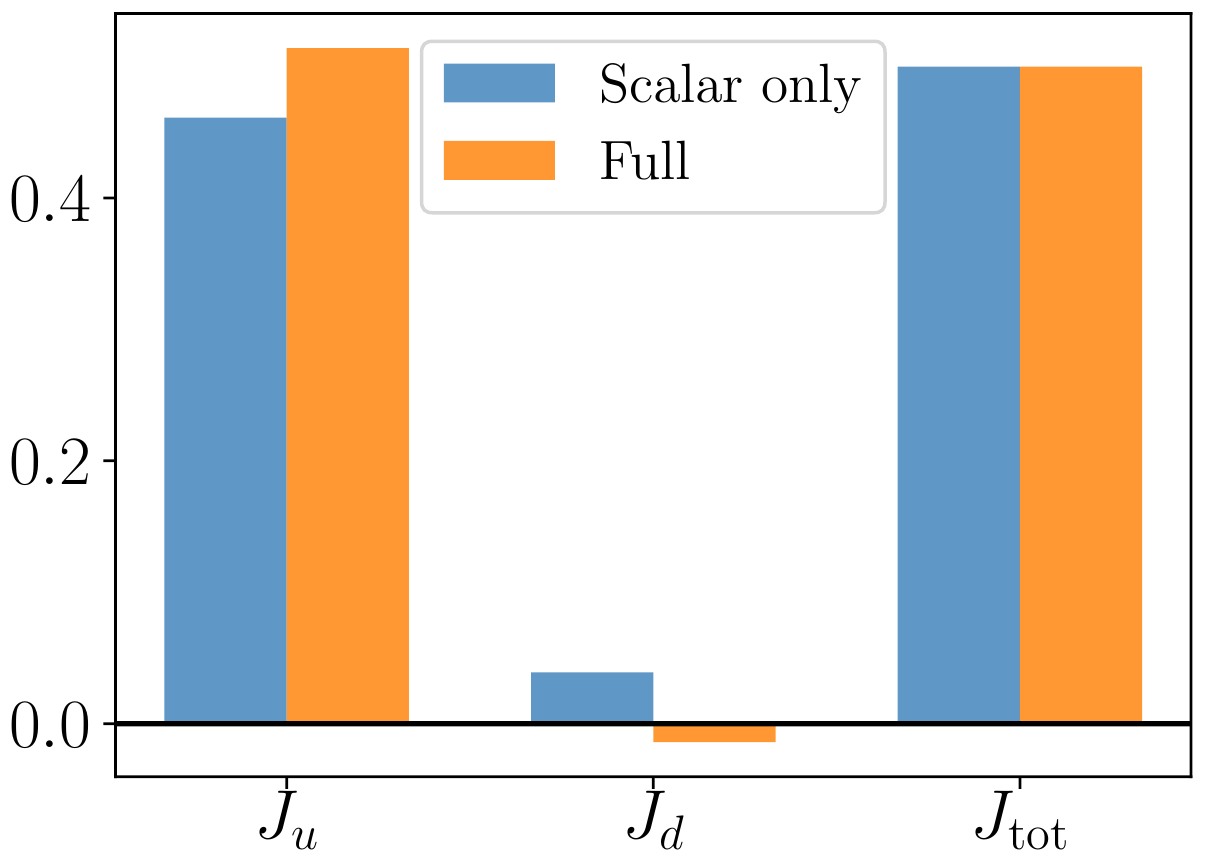}
  \caption{
    Spin decomposition at the model scale.
    The scalar diquark only model variant is contrasted with the full model
    containing both scalar and axial vector diquarks.
  }
  \label{fig:spin:decomp:njl}
\end{figure*}

For the total intrinsic spin contribution $S_{\rm tot} = S_u + S_d$
we find that scalar diquark configurations dominate,
even though the scalar diquark itself has no intrinsic spin.
For $L_{\rm tot}$ and $J_{\rm tot}$ the situation is more subtle because of
cancellations between different contributions.
However, we note that diquark transition diagrams cannot contribute to
conserved quantum numbers, since the scalar and axial vector diquark
configurations are effectively orthogonal states.
Moreover, the transition diagrams cannot contribute to any isoscalar quantities
such as $S$, $L$, or $C(0)$ because the transition itself is isovector
(namely, from an isovector to an isoscalar diquark, or vice-versa).
On the other hand, they can make a potentially large contribution
to isovector quantities such as $S_u-S_d$.
In fact, the transition diagram is responsible for nearly half of
our calculated value for $g_A$.
In this case, one can see that $g_A$ is an overestimate compared
to the experimental value of $g_A=1.2732(23)$~\cite{Tanabashi:2018oca}.
This discrepancy can be alleviated by the inclusion of meson cloud effects,
as done in Ref.~\cite{Cloet:2014rja} for the simpler calculations of proton
electromagnetic form factors.

\subsection{Proton spin decomposition}
The leading-twist proton GPDs allow us to obtain the Ji decomposition
of proton spin.
In particular, the quark total angular momentum $J_q$ can be broken up into
$S_q$ and $L_q$,
and the total gluon angular momentum $J_g$ can be obtained at an evolved scale
from the perturbatively generated gluon GPDs.
Since the Ji decomposition does not allow $J_g$ to be broken into spin and
orbital components,
we will use $S_{\mathrm{tot}}$ and $L_{\mathrm{tot}}$ to signify the
total \textsl{quark} spin and orbital angular momentum.

In Fig.~\ref{fig:spin:decomp:njl}, we compare the proton spin decomposition
at the model scale for both variants of our NJL model, that is,
one where the proton has only scalar diquark correlations and
the full model that also includes axial vector diquarks.
Remarkably, the total angular momentum carried by each quark flavor,
as well as the total quark spin and total quark orbital angular momentum
change very little when axial vector diquarks are introduced.
This may be attributed to the static approximation is used for the
quark-diquark interaction kernel, where orbital angular momentum is
generated by relativistic effects, in particular, by the presence of a
p-wave component in the quark wave function~\cite{Thomas:2008ga}.
Since the relativistic effects are about equally strong in both variants
of the model, $L_{\mathrm{tot}}$ and $S_{\mathrm{tot}}$ are about equal.

In the scalar-only model, $S_d=0$ because the down quark is present only
in the diquark, which does not allow a spin quantization axis to be identified.
Non-relativistically, one would have $L_d=J_d=0$ as well,
but the remaining quark in the proton carrying orbital angular momentum---since
it can exist in a p-wave state---implies that the diquark, and thus
the down quark, can carry orbital angular momentum as well.
The diagram breakdown for the full model in Tab.~\ref{tab:j:flavor}
indeed shows that $L_d$ and $J_d$ are non-zero because of the scalar diquark
diagram.

The flavor breakdown of $J$, $L$, and $S$ changes significantly when
axial vector diquarks are present, for two reasons.
The first---but more minor---reason is that the flavor breakdown within
axial diquark configurations differs from the scalar diquark case.
The effects of this are minimal however, and owe entirely to relativistic
effects.
Non-relativistically (and within the static approximation),
there is no orbital angular momentum,
and the axial diquark configuration has a spin wave function:
$\left|J_p^z = +\tfrac{1}{2}\right\rangle
  =
  \sqrt{\tfrac{2}{3}} \left|J^z_{\mathrm{dq}}=+1; J^z_q=-\tfrac{1}{2}\right>
  -
  \sqrt{\tfrac{1}{3}} \left|J^z_{\mathrm{dq}}=0; J^z_q=+\tfrac{1}{2}\right>
$
which when combined with the appropriate isospin recombination coefficients,
gives $J_d = S_d = 0$.
Indeed, even within the proper NJL model calculation,
we find the contributions to $J_d$ and $S_d$ from the axial diquark diagrams
are $0.006$ and $-0.021$, respectively.

\begin{table}[b]
  \setlength{\tabcolsep}{0.1em}
  \renewcommand{\arraystretch}{1.3}
  \caption{
    Flavor decomposition of the total, spin, and orbital angular momenta,
    where the results include contributions from both scalar
    and axial vector diquarks.
    Results are at the model scale of $Q^2 = 0.16$~GeV$^2$
    and the diquark in the brackets is the spectator.
  }
  \label{tab:j:flavor}
  \begin{tabular}{@{}lcccccc@{}}
    \toprule
    Diagram &
    $ J_u $ &
    $ J_d $ &
    $ S_u $ &
    $ S_d $ &
    $ L_u $ &
    $ L_d $ \\
    \hline
    Quark (scalar) &
    $\phantom{-}0.277 $ &
    $ 0 $ &
    $\phantom{-}0.287 $ &
    $ 0 $ &
    $-0.010 $ &
    $ 0 $ \\
    Scalar diquark &
    $\phantom{-}0.034 $ &
    $\phantom{-}0.034 $ &
    $ 0 $ &
    $ 0 $ &
    $\phantom{-}0.034 $ &
    $\phantom{-}0.034 $ \\
    Quark (axial) &
    $-0.013 $ &
    $-0.026 $ &
    $-0.022 $ &
    $-0.044 $ &
    $\phantom{-}0.009 $ &
    $\phantom{-}0.018 $ \\
    Axial diquark &
    $\phantom{-}0.163 $ &
    $\phantom{-}0.033 $ &
    $\phantom{-}0.114 $ &
    $\phantom{-}0.023 $ &
    $\phantom{-}0.049 $ &
    $\phantom{-}0.010 $ \\
    Transition diquark &
    $\phantom{-}0.054 $ &
    $-0.054 $ &
    $\phantom{-}0.188 $ &
    $-0.188 $ &
    $-0.134 $ &
    $\phantom{-}0.134 $ \\
    \hline
    Sum &
    $\phantom{-}0.514 $ &
    $-0.014 $ &
    $\phantom{-}0.567 $ &
    $-0.209 $ &
    $-0.053 $ &
    $\phantom{-}0.195 $ \\
    \bottomrule
  \end{tabular}
\end{table}

The most significant contributions to the change in flavor breakdown come from
transition diagrams.
Although $J$ is a conserved quantity, individual flavor contributions are not.
Moreover, individual flavor contributions are not isoscalar,
and in fact $J_u - J_d$ etc.\ are isovector,
meaning the transition diagram has the potential to make significant changes
to these differences.
In fact, as can be seen in Tab.~\ref{tab:j:flavor},
$S_d$ and $L_d$ are dominated by the transition diagram,
although this diagram makes a small contribution to $J_d$.

\begin{figure*}[tbp]
  \includegraphics[width=0.31\textwidth]{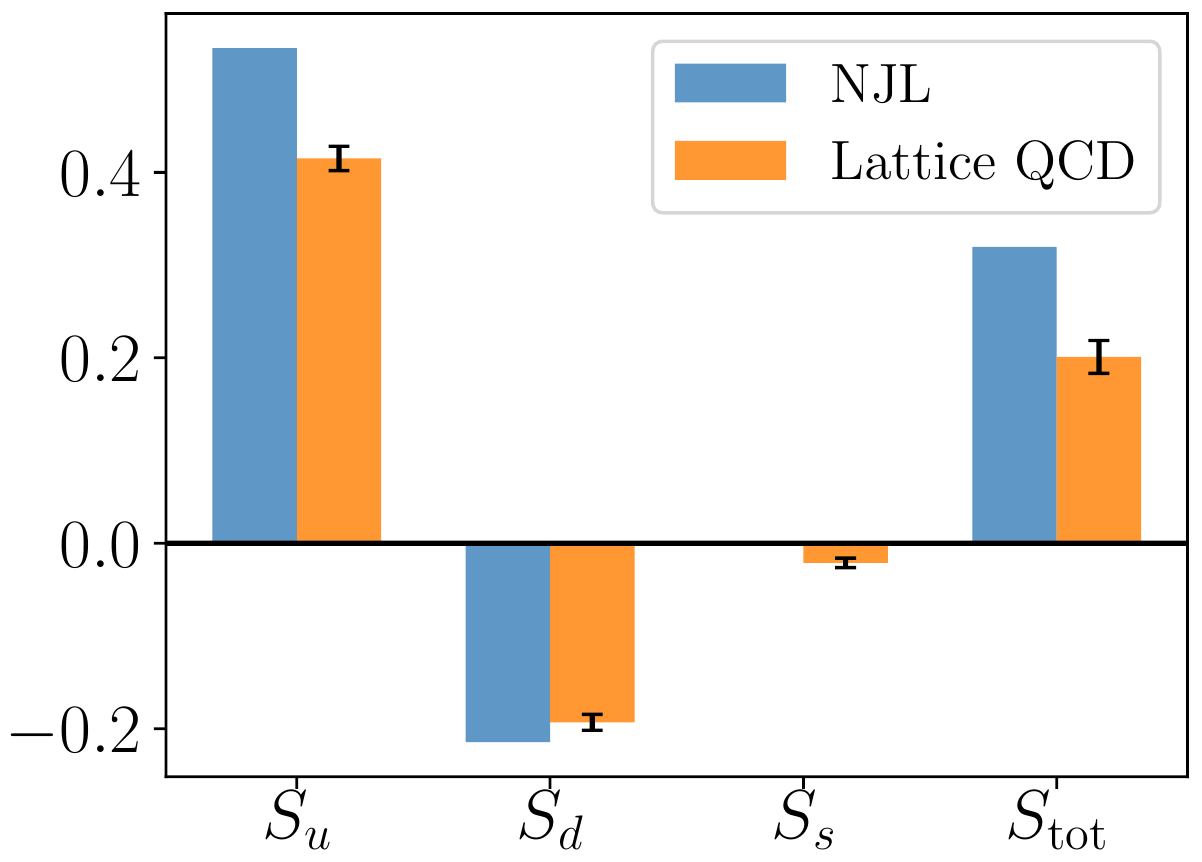} \hs*{3mm}
  \includegraphics[width=0.31\textwidth]{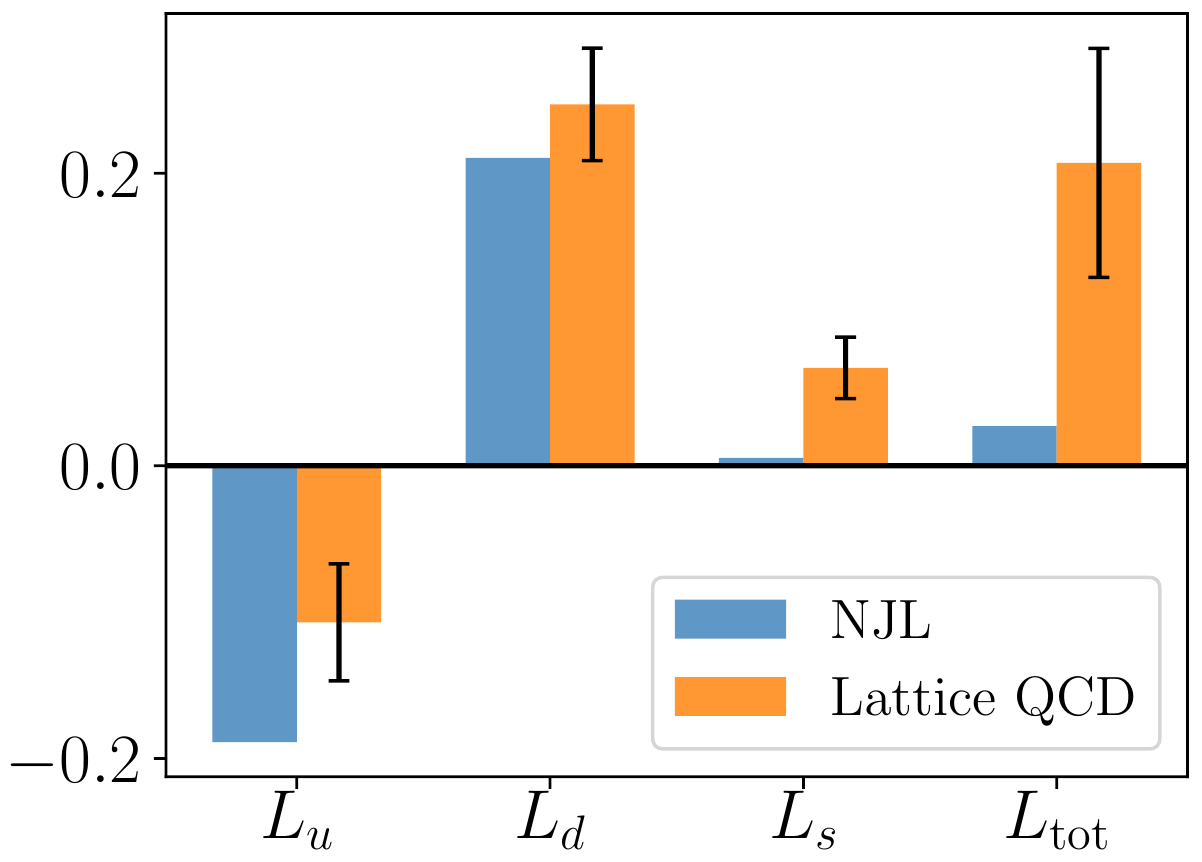} \hs*{3mm}
  \includegraphics[width=0.31\textwidth]{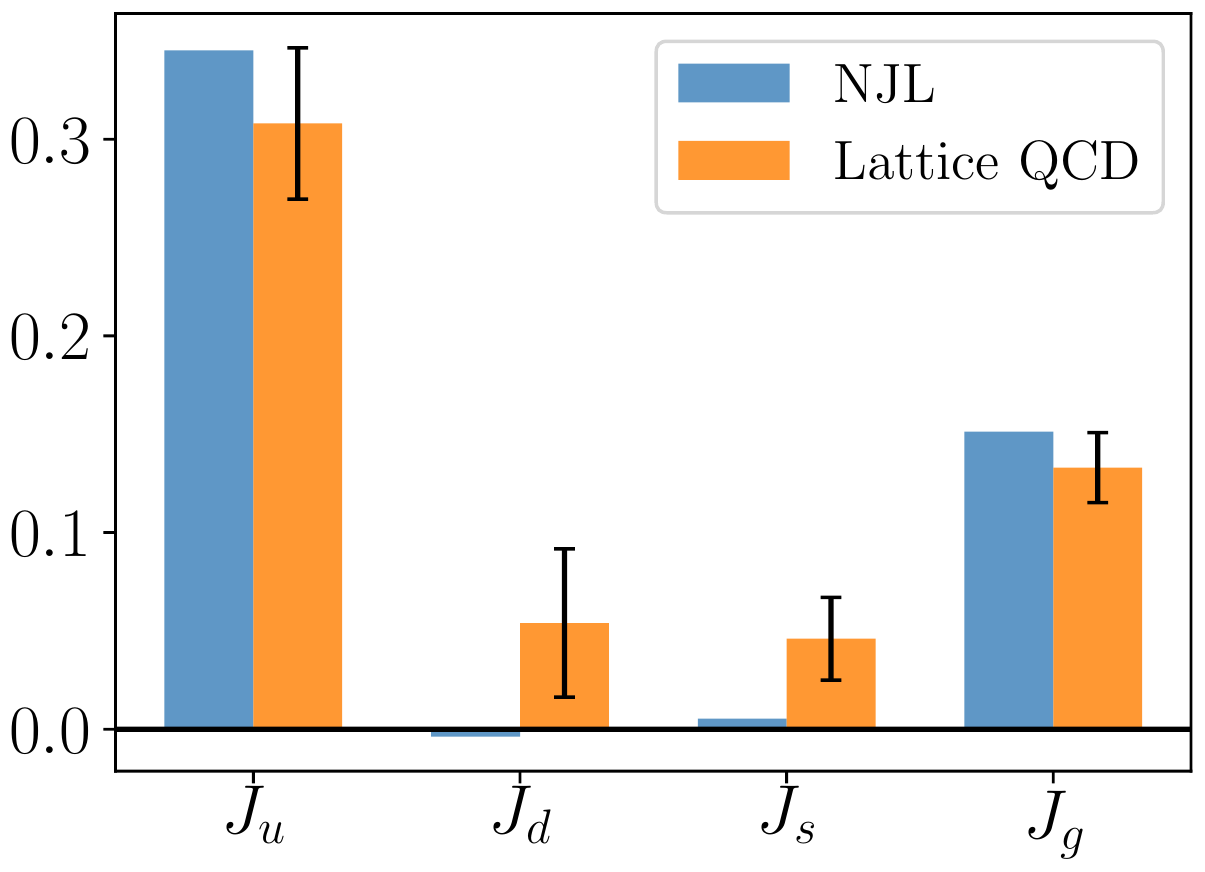}
  \caption{
    Comparison of NJL model spin decomposition to the lattice QCD
    results of Ref.~\cite{Alexandrou:2017oeh} at $Q^2=4$~GeV$^2$.
    The NJL results contain contributions from both scalar
    and axial vector diquarks.
  }
  \label{fig:spin:decomp:lattice}
\end{figure*}

Overall, $J_d$ comes out very close to zero in the scalar+axial model.
This can be seen as arising from cancellations.
On one hand, $S_d$ and $L_d$ end up being nearly equal and opposite
after contributions from all the diagrams have been summed.
On the other hand, in Tab.~\ref{tab:static:diagrams} one sees that
$A_d(0)$ and $B_d(0)$ are nearly equal and opposite after summing the diagrams.

Besides inter-model comparisons, it is worth comparing our flavor-separated
proton spin decomposition to the best available estimates for the true
proton spin decomposition.
Although experimental extractions for linear combinations of $J_u$ and $J_d$
exist from JLab~\cite{Mazouz:2007aa}
and HERMES~\cite{Ye:2006gza,Airapetian:2008aa},
these extractions are model-dependent and may not be instructive.
On the other hand, there exists a lattice QCD computation of the proton
spin decomposition at physical pion mass~\cite{Alexandrou:2017oeh}.

In Fig.~\ref{fig:spin:decomp:lattice},
we compare our results (with both diquark species present)
to the lattice results of Ref.~\cite{Alexandrou:2017oeh}
at a scale of $Q^2=4$~GeV$^2$.
One can observe mixed agreement with the lattice results.
Firstly, it's worth remarking that the broad
qualitative agreement on the sign and magnitude of
$J_u$, $J_d$, $L_u$, $L_d$, $S_u$, and $S_d$ is remarkable considering the
simplicity and minimalism of the NJL model.
This is suggestive that the spin decomposition of the proton is governed
to a large extent by three effects: its diquark content,
relativistic effects that can generate $L$,
and QCD evolution (which connects the model scale to the empirical scale).
Further intricacies (such as a meson cloud) could be
somewhat large but seem to be second-order effects.
Solving the Faddeev equation beyond the static approximation will
also have an impact, however since the spin decomposition is defined through
$t=0$ moments, the effects of exchange diagrams are expected to be small.

The agreement for $J_g$ is surprising,
since in our calculation this is generated purely by QCD evolution,
and is therefore suggestive of a small intrinsic gluon angular momentum.
Our calculations also tend to overestimate $S_q$ for the light quarks,
which could be rectified by the inclusion of a pion and kaon cloud,
which would also generate the missing intrinsic $S_s$ contributions.
Since $J_q$ agrees reasonably well with lattice,
corrections that decrease $S_q$ would at the same time need to increase $L_q$,
which is natural in the meson cloud picture because of their
p-wave couplings to the quarks or nucleon.

\section{SUMMARY AND OUTLOOK}
\label{sec:summary}

In this work, we calculated the helicity-dependent and helicity-independent
leading-twist proton GPDs in a confining version of the NJL model.
A quark-diquark approximation was used for the proton,
and two variants of the model were considered:
(1) a model with only isoscalar, Lorentz scalar diquarks; and
(2) a model also containing isovector, axial vector diquarks.
In both model variants, a flavor-separated spin decomposition was performed
for the proton,
and the presence of both diquark species was found to contribute significantly
to the flavor-separated spin decomposition,
but little to $S_{\mathrm{tot}}$ and $L_{\mathrm{tot}}$.
In particular, transition diagrams between the diquark species---which
can affect only isoscalar quantities, such as $S_u-S_d$---was responsible
for most of the difference between the models' spin decompositions.

The model variant with both diquarks present was found to have mixed agreement
with lattice results for the proton's spin decomposition.
The discrepancies are due primarily to the NJL model's overestimates of spin
and underestimates of orbital angular momentum,
along with the lack of strangeness content.
The former of these discrepancies can be resolved by the inclusion of a pion
cloud, and the latter with the inclusion of a kaon cloud.
These improvements warrant future work on the subject.


\begin{acknowledgments}
This work was supported by the U.S.~Department of Energy, Office of Science,
Office of Nuclear Physics, contract no.~DE-AC02-06CH11357,
and an LDRD initiative at Argonne National Laboratory
under Project~No.~2020-0020.
\end{acknowledgments}

\appendix
\section{Longitudinal-transverse separation for axial operators}
\label{sec:LT}

The non-local operator defining the helicity-dependent GPDs,
as well as the local axial current, are defined using matrix elements
of the operator $\bar{q}(x) \gamma_\mu \gamma_5 q(y)$,
with the spacetime points $x$ and $y$ determined by the application in question.
This operator notoriously does not correspond to a conserved current.
However, it is possible to break the operator into a ``transverse'' piece
that is conserved and a ``longitudinal'' piece that is not.
The breakdown is most clear in momentum space, where we define:
\begin{align}
  (\gamma_\mu\gamma_5)_\perp
  =
  \left(\gamma_\mu - \frac{\Delta_\mu\slashed{\Delta}}{\Delta^2}\right) \gamma_5
  \,, \qquad
  (\gamma_\mu\gamma_5)_\parallel
  =
  \frac{\Delta_\mu\slashed{\Delta}}{\Delta^2} \gamma_5
  \,,
\end{align}
with $\Delta$ being the momentum transfer to the target.
That $(\gamma_\mu\gamma_5)_\perp$ is transverse to $\Delta$ makes
$\bar{q}(x) (\gamma_\mu \gamma_5)_\perp q(x)$ a conserved local current.

Notably, the dynamics of
$\bar{q}(x) (\gamma_\mu \gamma_5)_\perp q(y)$
and
$\bar{q}(x) (\gamma_\mu \gamma_5)_\parallel q(y)$
completely decouple.
This means that the Bethe-Salpeter equations for the local currents
$\bar{q}(0) (\gamma_\mu \gamma_5)_\perp q(0)$
and
$\bar{q}(0) (\gamma_\mu \gamma_5)_\parallel q(0)$
decouple from each other,
and also that the BSEs for the leading-twist non-local correlators
\begin{align}
  &A_{\perp,\parallel}^q
  =
  \frac{1}{2}
  \int \frac{\mathrm{d}z}{2\pi}
  e^{ix(Pn)\kappa}\no \\
&\hs*{3mm}\times
  \left\langle p'\lambda' \middle|
  \bar{q}\left(-\frac{nz}{2}\right)
  (\slashed{n} \gamma_5)_{\perp,\parallel}
  \left[ -\frac{nz}{2}, \frac{nz}{2} \right]
  q\left(\frac{nz}{2}\right)
  \middle| p\lambda\right\rangle,&%
\end{align}
decouple from each other.

The Lorentz decompositions of the transverse and longitudinal
components of the helicity-dependent correlator can be written,
for a spin-half particle, as:
\begin{align}
  A_{\perp,\lambda'\lambda}^q
  &=
  \bar{u}(p',\lambda') (\slashed{n}\gamma_5)_\perp u(p,\lambda)
  \, \widetilde{H}(x,\xi,t),
  \\
  A_{\parallel,\lambda'\lambda}^q
  &=
  \bar{u}(p',\lambda') (\slashed{n}\gamma_5)_\parallel u(p,\lambda)
  \,
  \left[ \widetilde{H}(x,\xi,t)
  - \tau \widetilde{E}(x,\xi,t) \right] \no \\
  &\equiv
  \bar{u}(p',\lambda') (\slashed{n}\gamma_5)_\parallel u(p,\lambda)
  \,
  \widetilde{E}_\parallel(x,\xi,t)
  \,.
\end{align}
Comparing to Eq.~(\ref{eqn:gpd:quark}),
we observe that the pion pole can contribute only to the longitudinal
component of the correlator.
This additionally means that the pion pole will not be present in
$\widetilde{H}^q(x,\xi,t)$---nor $G_A(t)$---of the proton.

For an on-shell spin-one particle,
the longitudinal-transverse separation can be written:
\begin{subequations}
  \begin{align}
    A_{\perp,\lambda'\lambda}^q
    &=
    - i
    \Bigg(
      \frac{\epsilon_{n\Delta P \rho}}{\Delta^2}
      \frac{
        \varepsilon^\rho(\varepsilon'^{*}\Delta)
        - \varepsilon'^{*\rho}(\varepsilon\Delta)
      }{(Pn)} \no \\
&\hs*{17mm}
      +
      \frac{(P\Delta)}{\Delta^2}
      \frac{\epsilon_{n\varepsilon\varepsilon'^*\Delta}}{(Pn)}
      \Bigg)
    \left[
      - \widetilde{H}_{1}^q
      + \frac{\Delta^2}{M_a^2} \widetilde{H}_{2}^q
      \right]
    \notag \\ &
    -
    \frac{i\epsilon_{n\Delta P \rho}}{M_a^2}
    \frac{
      \varepsilon^\rho(\varepsilon'^{*}\Delta)
      + \varepsilon'^{*\rho}(\varepsilon\Delta)
    }{(Pn)} \, \widetilde{H}_{3}^q \no \\
&
    +
    \frac{i\epsilon_{n\Delta P \rho}}{2(Pn)}
    \frac{
      \varepsilon^\rho(\varepsilon'^{*}n)
      + \varepsilon'^{*\rho}(\varepsilon n)
    }{(Pn)} \, \widetilde{H}_{4}^q,
    \label{eqn:spin1:LT:T}  \displaybreak[0]
    \\
    A_{\parallel,\lambda'\lambda}^q
    &=
    \frac{(n\Delta)}{\Delta^2}
    \frac{i\epsilon^{\Delta\varepsilon\varepsilon'^*P}}{(Pn)}
    \, \widetilde{H}_{1}^q
    \label{eqn:spin1:LT:L}
    \,.
  \end{align}
  \label{eqn:spin1:LT}%
\end{subequations}
This breakdown agrees exactly with the standard breakdown
in Ref.~\cite{Berger:2001zb} for on-shell particles,
through use of the Schouten identity result:
\begin{align}
  (n&\Delta)
  \epsilon^{\Delta \varepsilon\varepsilon'^*P}
  =
  \Delta^2
  \epsilon^{n \varepsilon \varepsilon'^* P} \no \\
&\hs*{0mm}
  -
  \epsilon^{n\Delta P \sigma}
  \Big(
    \varepsilon_\sigma (\varepsilon'^*\Delta)
    -
    \varepsilon'^*_\sigma (\varepsilon\Delta)
    \Big)
  -
  (\Delta P) \epsilon^{n\varepsilon\varepsilon'^*\Delta}\,,
  \label{eqn:schouten}
\end{align}
and the on-shell relation $(\Delta P)=0$,
as well as use of the identities
$(\varepsilon \Delta) = 2 (\varepsilon P)$
and $(\varepsilon'^* \Delta) = -2 (\varepsilon^* P)$.

For an off-shell particle, $(\Delta P)\neq0$ means the equivalence between
the decompositions no longer holds.
Crucially,
the decompositions differ by a transverse structure that multiplies a
longitudinal GPD.
This means using the standard decomposition for an off-shell spin-one
particle will introduce unphysical pion poles into transverse quantities,
such as $G_A(t)$ of the proton.
Therefore, the alternative decomposition suggested in
Eq.~(\ref{eqn:spin1:LT})---which indeed does not produce unphysical
pion poles in the proton's axial form factor---is preferred for the off-shell
spin-one correlator.

One last crucial aspect of Eq.~(\ref{eqn:spin1:LT}) worth remarking on is the
explicit inclusion of a term proportional to $(P\Delta)$.
For an on-shell spin-one particle, this term is zero and is not important.
For an off-shell particle, however, it is necessary for the axial correlator
$A^q_{\lambda'\lambda}$ to be analytic at $t=0$.
Neither the Lorentz structure multiplying $\widetilde{H}^q_1$ in
Eq.~(\ref{eqn:spin1:LT:L})
nor the structure multiplying
$-\widetilde{H}_1^q + \frac{t}{M_a^2}\, \widetilde{H}_2^q$
in Eq.~(\ref{eqn:spin1:LT:T})
has a well-defined forward limit;
if one writes $\Delta^\mu = \sqrt{-t}\, e^\mu$,
with $e^\mu$ an arbitrary spacelike unit vector,
then the $t\rightarrow0^-$ limit depends on $e^\mu$,
which is unphysical.
However, by virtue of the Schouten identity Eq.~(\ref{eqn:schouten})
and the presence of the $(P\Delta)$ term in Eq.~(\ref{eqn:spin1:LT:T}),
the total axial correlator
$A^q_{\parallel,\lambda'\lambda} + A^q_{\perp,\lambda'\lambda}$
does have a well-defined $t=0$ limit, even when $(P\Delta)\neq0$.


\bibliography{main.bib}

\end{document}